\documentclass{article}

\usepackage{algorithmic}
\usepackage{algorithm}
\usepackage{enumitem}
\usepackage{arxiv}

\usepackage{amsmath}
\usepackage{amssymb}
\usepackage{array}

\usepackage{authblk}
\usepackage{float}

\usepackage{footnote}

\usepackage[utf8]{inputenc} % allow utf-8 input
\usepackage[T1]{fontenc}    % use 8-bit T1 fonts
\usepackage{hyperref}       % hyperlinks
\usepackage{url}            % simple URL typesetting
\usepackage{booktabs}       % professional-quality tables
\usepackage{amsfonts}       % blackboard math symbols
\usepackage{nicefrac}       % compact symbols for 1/2, etc.
\usepackage{microtype}      % microtypography
\usepackage{lipsum}
\usepackage{graphicx}
\graphicspath{ {./images/} }

\title{DIVERSE: Bayesian Data IntegratiVE learning for precise drug ResponSE prediction}

\author[1]{ Bet\"{u}l~G\"{u}ven\c{c} Paltun}
\author[1, 2]{ Samuel~Kaski}
\author[1, 3]{ Hiroshi~Mamitsuka}

\affil[1]{\footnotesize Department of Computer Science, Aalto University, Finland}
\affil[2]{\footnotesize Department of Computer Science, University of Manchester, UK}
\affil[3]{\footnotesize Bioinformatics Center, Institute for Chemical Research, Kyoto University, Japan}

\begin{document}
\maketitle
\begin{abstract}
Detecting predictive biomarkers from multi-omics data is important for precision medicine, to improve diagnostics of complex diseases and for better treatments. This needs substantial experimental efforts that are made difficult by the heterogeneity of cell lines and huge cost. An effective solution is to build a computational model over the diverse omics data, including genomic, molecular, and environmental information. However, choosing informative and reliable data sources from among the different types of data is a challenging problem. We propose DIVERSE, a framework of Bayesian importance-weighted tri- and bi-matrix factorization(DIVERSE3 or DIVERSE2) to predict drug responses from data of cell lines, drugs, and gene interactions. DIVERSE integrates the data sources systematically, in a step-wise manner, examining the importance of each added data set in turn. More specifically, we sequentially integrate five different data sets, which have not all been combined in earlier bioinformatic methods for predicting drug responses. Empirical experiments show that DIVERSE clearly outperformed five other methods including three state-of-the-art approaches, under cross-validation, particularly in out-of-matrix prediction, which is closer to the setting of real use cases and more challenging than simpler in-matrix prediction. Additionally, case studies for discovering new drugs further confirmed the performance advantage of DIVERSE.
\end{abstract}

% keywords can be removed

\section{Introduction}
Identification of predictive biomarkers for drug sensitivity plays a significant role for assigning the most effective treatments to patients with complex diseases such as cancer \cite{la2011predictive}. However, it is impracticable to clinically assess each patient’s response to disease due to the large population. Patients of the same cancer type may differ in their responses to a specific medical therapy because of the large genetic diversity of cancer \cite{zhang2018hybrid}. Personalized medicine provides an understanding of cancer cell lines at the molecular level and recommends individualized therapies to patients that allow high efficacy in different cancer types by measuring drug responses \cite{guvencc2019improving}. 

The research is most often done with cell lines which, even though much simpler than real patients, are already complex enough and require multiple data sets to characterize sufficiently for prediction. Since cancer cell lines show distinct characteristics caused by a multitude of factors, including genetic mutations, molecular interactions, and environmental sources, complicates the discovery of predictive biomarkers. Fortunately, recent high-throughput technologies have generated a considerable amount of biological data from different viewpoints. This diverse data could allow precise computational prediction of drug sensitivity of cancer cell lines based on molecular interactions, genomic features, and chemical structures. However, although large-scale data have been generated for drug response prediction, many machine learning methods have failed to achieve good performance for multiple heterogeneous data sources, because these methods have been designed for only a single type of data. Thus a challenging task is to build precise prediction models on diverse data, coming from different sources, which are difficult to compare. In fact, data integration has to overcome several obvious problems, such as different data sizes, complexity, and noisiness. However, more importantly, data-integrative machine learning methods need to decide which information is useful to be incorporated and how significant the information is for the prediction task. This is the most critical problem to be addressed for machine learning models with diverse multi-omics data. For this problem, we propose DIVERSE, a framework to efficiently integrate scientifically diverse data, i.e. genomic, chemical and molecular interaction information, to predict missing drug responses of cancer cell lines. The three key points of DIVERSE are:

i) DIVERSE integrates five biologically different data sets: drug similarity, gene expression, protein-protein interaction, drug-target interaction and cell line-drug interaction.
To the best of our knowledge, this is the largest number of heterogeneous data sources combined for drug response prediction so far. No competing bioinformatics methods can integrate the same types of data sets.

ii) It does not allow any of the five different data sets dominate the prediction. DIVERSE adds one data set by one in a systematic and step-wise manner.

iii) It is methodologically flexible. Most existing studies ignore uncertainty, and hence cannot accept missing values. Second, in general, integrating different data sets makes it harder to obtain the correct rank of given data or matrices. DIVERSE solves these two practically important problems by using a Bayesian setting. 

We empirically validated the performance of DIVERSE, comparing with five other methods, including three state-of-the-art methods, under 5x5-fold cross-validation. Experimental results indicate that DIVERSE significantly outperformed all compared methods in both mean-squared error (MSE) and Spearman correlation coefficient (Sc), particularly for out-of-matrix prediction, which is a real-world setting and much harder than in-matrix prediction. Results clearly show the performance advantage of DIVERSE over the current methods for predicting drug responses. Also, the results indicate that the MSE and Sc of DIVERSE were smoothly improved by the step-wise addition of each data set. These advantages of DIVERSE were confirmed by several case studies.

\section{Related work}
The promise of personalized medicine has been a theme in clinical discussion for some time, and researchers have developed a variety of computational methods. Some state-of-the-art algorithms focusing on drug response prediction include elastic net \cite{barretina2012cancer}, support vector machines \cite{menden2013machine}, kernel ridge regression (KRR) \cite{murphy2012machine}, random forest and neural networks \cite{costello2014community}. However, most of the traditional approaches underestimate the complexity of cancer caused by a number of environmental factors, genetic mutations, and somatic alterations in drug response prediction. A variety of studies have been taking into account the complex relationships between cell lines, chemical structures, and genomic alterations to identify predictive biomarkers \cite{guvencc2019improving}.

The advantage of incorporating heterogeneous information for drug response prediction analysis has been highlighted in recent studies \cite{liu2018anti, guan2019anticancer}. Ammad-ud din et al. proposed in their cwKBMF that utilizing genomic data increases predictive performance, and incorporating prior biological knowledge enhances it even further \cite{ammad2016drug}. SRMF \cite{wang2017improved} was proposed as a matrix factorization method to simultaneously incorporate drug and cell line similarity information for drug response prediction. 
Multiple non-negative matrix factorization (MultiNMF) models have been designed for integrating data sets by sharing one of the factor matrices, details can be found in the study by \cite{fujita2018biomarker}. DrugCellNet \cite{zhang2015predicting} assumes that the response of a known drug in a new cell line is a weighted combination of the responses of the neighboring cell lines. \cite{stanfield2017drug} focused on solving the “small n, large p” problem when the number of genes is larger than samples, through integrating cell line and drug information with a PPI network by utilizing functional links. HNMDRP was proposed by \cite{zhang2018novel} as a classification problem of whether that the drug is whether sensitive or resistant, based on data on gene expression profiles from cell lines, drug chemical structure features, drug-target interactions, and PPIs. Even though HNMDRP utilizes enough data, it cannot predict unseen (new) drugs and cell lines. Recently, \cite{cichonska2018learning} developed a time- and memory-efficient learning method with multiple pairwise kernels that can integrate various types of biological data sources, which, however, requires that the data comes in the form of observations for pairs of entities, such that data only includes drug and cell line information. Another network-based drug response prediction method MOLI \cite{sharifi2019moli}, was recently proposed and built on deep neural networks for feature selection. MOLI learns the features of each data item, and then concatenates them for predicting drug response values. Detailed comparisons of recent machine learning models in data-integrative drug response prediction can be found in a recent review by \cite{guvencc2019improving}.

DIVERSE has three unique aspects to address the limitations of existing models and to develop more robust models. That is, DIVERSE can 1) integrate several entities such as drugs, cell lines, and genes to predict missing entries and unseen drugs. For example, many methods have to use cell line similarity information transformed from gene expression data since they can not integrate gene-related information; 2) use a computationally feasible method to integrate data sets because when a new data set is combined, DIVERSE can use the same entity-specific factors without additional computation; 3) use non-negativity constraints in a Bayesian setting to reduce overfitting on noisy data, sustaining better interpretation after factorization.

 \begin{figure}[!t]
	\centering
	\includegraphics[width=0.3\linewidth]{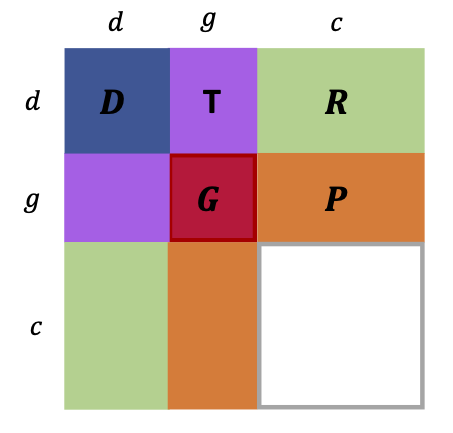}
	\caption{Conceptual integration configuration of the multiple data from three types of entities; $d$, $g$ and $c$ denote drugs, genes and cell lines respectively.}
	\label{fig:all_data}
\end{figure}

\section{Methods}
DIVERSE allows predicting drug responses of cancer cell lines by incorporating information from heterogeneous data sets. DIVERSE consists of two key elements: 1) Section 3.1.2--4: Bayesian non-negative matrix factorization that is used to determine latent factors of data sets, including data describing relations between drugs, cell lines, and genes. 2) Section 3.1.5--6: hybrid matrix factorization model to simultaneously integrate heterogeneous data sets. This combination of methods is new for predicting drug sensitivity.

\subsection{Prediction model}

\subsubsection{Prediction problem}

The goal of this work is to predict missing entries of a drug response matrix given the other matrices. This problem consists of two different tasks. First, we predict an unknown value of a pair of a drug and a cell line, for a drug for which other values are already given (observed). Second, we predict all responses of an unseen (new) drug which has no observed values in the matrix yet. Drug response data consists of IC50 values that give the effectiveness of drugs on different cell lines. Additional given inputs are drug  similarity, gene expression, protein-protein interaction, drug-target interaction and cell  line-drug  interaction data sources. Details of the data will be described in Section \ref{sec:data}.

 \begin{table}[!t]
\renewcommand{\arraystretch}{1.3}
 \caption{The List of Symbols and Notations Used in This Paper}
\label{Tab:notations}
 \centering
\addtolength{\tabcolsep}{-7pt}
\begin{tabular}{lc}
\hline 
Symbol &  Description \\ \hline
$\mathbf{R}$ &  \text Drug response matrix (main input), $\mathbf{R} \in \mathbb{R}^{N \times M}$ \\ 
$\mathbf{D}$ & \text{Drug similarity matrix}, $\mathbf{D} \in \mathbb{R}^{N \times N}$\\ 
$\mathbf{P}$ & \text Gene expression matrix, $\mathbf{P} \in \mathbb{R}^{M \times L}$ \\ 
$\mathbf{G}$ & \text Protein-protein interaction matrix, $\mathbf{G} \in \mathbb{R}^{L \times L}$\\ 
$\mathbf{T}$ & \text Drug-target interaction matrix, $\mathbf{T} \in \mathbb{R}^{L \times N}$\\ 
$\mathbf{U}$ &\text Low-rank representation of drugs, $\mathbf{U} \in \mathbb{R}^{N \times K_{d}}$ \\ 
$\mathbf{V}$ & \text Low-rank representation of cell lines, $\mathbf{V} \in \mathbb{R}^{M \times K_{c}}$ \\
$\mathbf{H}$ & \text Low-rank representation of genes, $\mathbf{H} \in \mathbb{R}^{L \times K_{g}}$ \\ 
$\mathbf{S_{r}}$ & \text Low-rank relation matrix of drugs and cell lines, $\mathbf{S_{r}} \in \mathbb{R}^{K_{d} \times K_{c}}$ \\ 
$\mathbf{S_{p}}$ & \text Low-rank relation matrix of cell lines and genes, $\mathbf{S_{p}} \in \mathbb{R}^{K_{c} \times K_{g}}$ \\
$\mathbf{S_{d}}$ & \text Low-rank similarity matrix of drugs, $\mathbf{S_{d}} \in \mathbb{R}^{K_{d} \times K_{d}}$ \\ 
$\mathbf{S_{g}}$ & \text Low-rank similarity matrix of genes, $\mathbf{S_{g}} \in \mathbb{R}^{K_{g} \times K_{g}}$ \\
$\mathbf{S_{t}}$ & \text Low-rank relation matrix of drugs and genes, $\mathbf{S_{t}} \in \mathbb{R}^{K_{g} \times K_{d}}$ \\ 
$\mathbf{\lambda_{k}}$ & \text Set of prior parameters of latent factors, $\lambda_{k} = \{\lambda_{k_{d}},\lambda_{k_{c}}, \lambda_{k_{g}}\}$  \\ 
$w$ &  Set of importance weights; $w=\{w_{r}, w_{p}, w_{d}, w_{t}, w_{g}\}$  \\
$\tau$ &  Set of noise parameters; $\tau=\{\tau_{r}, \tau_{p}, \tau_{d}, \tau_{t}, \tau_{g}\}$ for all data \\
$M$ &  Main block includes main prediction data \\
$F$ &  Feature block includes feature matrices \\
$S$ &  Similarity block includes similarity matrices\\ 
\hline
\end{tabular}
\end{table}

\subsubsection{Notation}

The main input is the drug response matrix $\mathbf{R} \in \mathbb{R}^{N\times M}$, in which rows correspond to drugs and columns to cell lines. Each entry in $\mathbf{R}$ is the response value of a single drug in a certain cell line. In order to represent associations between cell lines and genes, we use $\mathbf{P} \in \mathbb{R}^{M\times L}$. The chemical similarity matrix of drugs is encoded as $\mathbf{D} \in \mathbb{R}^{N\times N}$. $\mathbf{T} \in \mathbb{R}^{L\times N}$ is the drug-target interaction matrix, where $\mathbf{T}_{ij}$ 1 if there is interaction between drug $i$ and gene $j$. Similarly, protein-protein interaction (PPI) matrix $\mathbf{G} \in \mathbb{R}^{L\times L}$ represents the functional relations between proteins. Here matrices are denoted by capital letters. The detailed interaction between matrices can be seen in Fig. \ref{fig:all_data}. Table~\ref{Tab:notations} shows the list of notations used throughout this paper.

\subsubsection{Non-negative matrix tri-factorization}

The drug response matrix $\mathbf{R}$ can be mapped to a non-negative low-dimensional latent factor space and regarded as the product of three matrices as follows: 
\begin{equation}
\mathbf{R} \approx  \mathbf{U S_{r} V}^{\mathsf{T}}. \quad \textrm{Here} \quad \mathbf{U}, \mathbf{V}, \mathbf{S_{r}} \in \mathbb{R}^{+}, \textrm{and}
\label{eq:R}
\end{equation}
$\mathbf{U} \in \mathbb{R}^{N \times K_{d}}$, $\mathbf{V} \in \mathbb{R}^{M \times K_{c}}$ describe the relationship of the latent factors to drugs and cell lines, respectively. The $\mathbf{S_{r}} \in \mathbb{R}^{K_{d} \times K_{c}}$ defines the latent relation between drugs and cell lines.

We use a probabilistic approach to formulate the factorization, which allows us to handle missing values efficiently. We assume a priori that each relation is drawn from Gaussian distribution with precision $\tau$. The likelihood function for the observed data is: 
\begin{align} 
 p(\mathbf{R}|\mathbf{U},\mathbf{S_{r}}, \mathbf{V},\tau_r) & = \prod_{i,j} \mathcal{N}(\mathbf{R}_{ij};\mathbf{U}_{i}\cdot \mathbf{ S_{r}} \cdot \mathbf{V}_{j}^{\mathsf{T}}, \tau_{r}^{-1}).
\label{eq:R2}
\end{align}
We choose priors that allow us to constrain latent matrices to be non-negative and permit an efficient inference procedure. Thus, the priors for the latent matrices are  chosen to be exponentially distributed with scales $\lambda_{k_{d}}$, and $\lambda_{k_{c}}$,
\begin{align} 
\mathbf{U}_{ik_{d}} \sim \textrm{Exp}(\lambda_{k_{d}}), \quad \mathbf{V}_{ik_{c}} \sim \textrm{Exp}(\lambda_{k_{c}}),  \quad \mathbf{S_{r}} \sim \textrm{Exp}(\lambda_{s_{r}})
\label{eq:priors}
\end{align}
where $k_{d} \in \{1,\dots,K_{d}\}$ and $k_{c}\in \{1,\dots,K_{c}\}$. The model is formulated with conjugate priors where noise variance is chosen as gamma distribution with shape $\alpha_{r}$ and scale $\beta_{r}$,
\begin{equation}
\tau_{r} \sim \mathcal{G}(\tau_{r};\alpha_{r},\beta_{r}).
\label{eq:tau}
\end{equation}

\subsubsection{Selecting the rank}

Bayesian settings help to seek the exact rank automatically in contrast to traditional matrix factorization methods by having a prior, and when integrating out parameters, the model ends up realizing that some components have zero contribution to the result. Instead of performing model selection to find the number of ranks for latent matrices, we define an upper bound, and the model determines the correct number of components. We define hyperpriors over prior parameters which are shared by columns of matrices to perform automatic model selection;
\begin{align} 
\lambda_{k_{d}} \sim \mathcal{G}(\lambda_{k_{d}};\alpha_{k_{d}},\beta_{k_{d}}), \quad  \lambda_{k_{c}} \sim \mathcal{G}(\lambda_{k_{c}};\alpha_{k_{c}},\beta_{k_{c}})
\label{eq:rank}
\end{align}
as used by \cite{tan2012automatic}. If prior has a low value, the entire column will be activated or eliminated if prior has a high value.

\subsubsection{Inference}

 \begin{algorithm}[!t]
 \caption{Gibbs sampling algorithm for drug response prediction}
 \label{pseudogibbs}
 \begin{algorithmic}[1]
 \renewcommand{\algorithmicrequire}{\textbf{Input:}}
 \renewcommand{\algorithmicensure}{\textbf{Output:}}
 \REQUIRE Drug response matrix $\mathbf{R}$
 \ENSURE  Approximated drug response matrix 
  \STATE \textit{Initialize model parameters}: $\mathbf{U}_{0}, \mathbf{S_{r}}_{0}, \mathbf{V}_{0},\lambda_{k_{d}^{0}}, \lambda_{s_{r}^{0}}, \lambda_{k_{c}^{0}}$
  \FOR {each iteration: i = 1, ..., T}
  \STATE Sample model hyperparameters:
  \STATE \hspace{0.05cm} $\lambda_{k_{d}}^{t}\sim p(\lambda_{k_{d}}|\mathbf{U}^{t},\lambda_{k_{d}^{0}})$
  \STATE \hspace{0.05cm} $\lambda_{s_{r}}^{t} \sim p(\lambda_{s_{r}}|\mathbf{S_{r}}^{t},\lambda_{s_{r}^{0}}))$
   \STATE \hspace{0.05cm} $\lambda_{k_{c}}^{t} \sim p(\lambda_{k_{c}}| \mathbf{V}^{t},\lambda_{k_{c}^{0}}))$
    \FOR{each drug, i = 1,...,N}
    \STATE \hspace{0.1cm} $\mathbf{U}_{i}^{t+1} \sim p(\mathbf{U}_{i}|\mathbf{R},\mathbf{S}^{t}, \mathbf{V}^{t},\lambda_{k_{d}}^{t})$
    \ENDFOR
    \FOR{each relation, $k=1, ..., K_{d}$}
    \STATE \hspace{0.1cm} $\mathbf{S_{r}}_{k}^{t+1} \sim p(\mathbf{S_{r}}_{k}|\mathbf{R},\mathbf{U}^{t}, \mathbf{V}^{t},\lambda_{s_{r}}^{t})$
    \ENDFOR
    \FOR{each cell lines, j = 1,...,M}
    \STATE \hspace{0.1cm} $\mathbf{V}_{j}^{t+1} \sim p(\mathbf{V}_{i}|\mathbf{R},\mathbf{S}^{t}, \mathbf{U}^{t},\lambda_{k_{c}}^{t})$
    \ENDFOR
  \ENDFOR
  \end{algorithmic} 
 \end{algorithm}

Given the observed measurements of cell lines, drugs, and genomic features, the posterior distribution of the model parameters is computed via the Bayes theorem. Since the model has been formulated with conjugate priors, Gibbs sampling can be conveniently used to sample new values for each parameter from their conditional distribution of given data and the current values of the other parameters. Derivation of conditional distributions from posterior distribution is straightforward due to using conjugate priors. The detailed iterative process of Gibbs sampling is summarized in Algorithm \ref{pseudogibbs}.

\subsubsection{Hybrid matrix factorization}

\enlargethispage{\baselineskip}

The purpose of our model is to predict the missing responses of drugs to given cell lines and unseen drugs for given multiple cancer cell lines by incorporating prior information. In order to infer drug responses of cancer cell lines and improve the accuracy, we apply Bayesian hybrid matrix factorization (HMF) to integrate several data sets concurrently as side information \cite{pmlr-v54-brouwer17a}.

HMF considers heterogeneous integration over three types of data blocks, each being a set of matrices: 1) main block $\mathcal{M}$ has the main matrices to be considered, 2) similarity block $\mathcal{S}$ has similarity matrices, and 3) feature block $\mathcal{F}$ has matrices, each relating two entities, while these matrices are not in $\mathcal{M}$.
HMF is suitable for our problem setting of integrating multiple data matrices, at the same time by sharing factors between data sets. Additionally, sharing latent matrices can be effectively used for data integration and improve the factorization \cite{zhang2005two}.

Let us show an example of HMF, with main block $\mathcal{M} = \{ \mathbf{R}, \mathbf{P} \}$ and similarity block $\mathcal{S} = \{ \mathbf{D}, \mathbf{G} \}$. We then simultaneously factorize these four matrices, where each matrix is factorized into a product of three non-negative low-dimensional matrices. Factorization of $\mathbf{R}$ is already given in (\ref{eq:R}), and then the rest three matrices, $\mathbf{P}$, $\mathbf{D}$, and $\mathbf{G}$, can be factorized as follows:
\begin{align} 
\mathbf{P} \approx  \mathbf{V S_{p} H}^{\mathsf{T}}, \quad \mathbf{D} \approx  \mathbf{U S_{d} U}^{\mathsf{T}}, \quad \mathbf{G} \approx  \mathbf{H S_{g} H}^{\mathsf{T}},
\label{eq:all_factors}
\end{align}
with the additional constraint that all latent factors are non-negative. Thus you can easily see that four input matrices can be factorized into only three latent factors, $\mathbf{V}$, $\mathbf{H}$, and $\mathbf{U}$ except $\mathbf{S}_*$.

\begin{figure*}[!t]
\centering
\includegraphics[width=\textwidth]{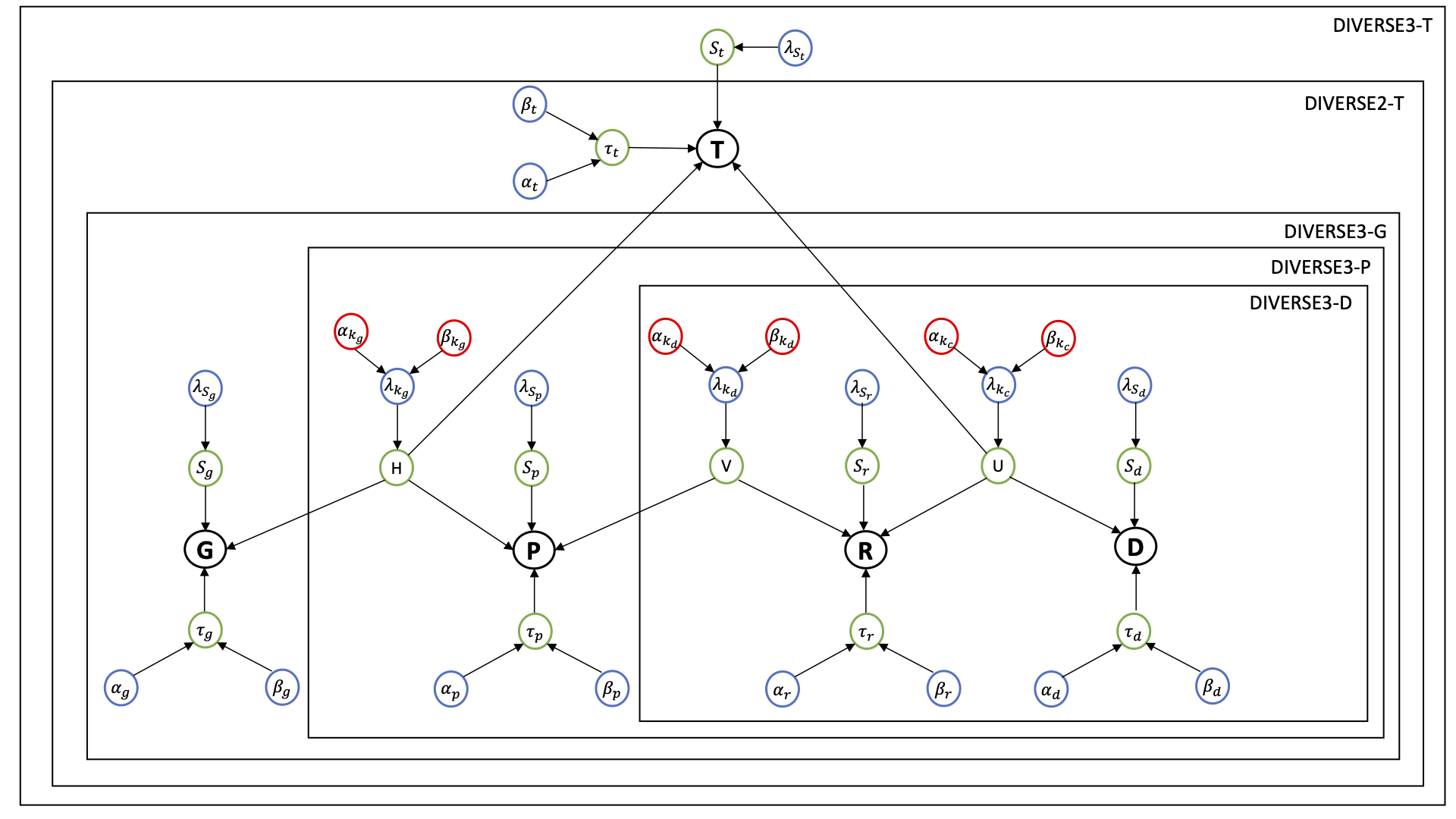}
\caption{ Graphical illustration of the DIVERSE method used for prediction of drug responses. The figure demonstrates given matrices, shared latent variables with their priors any hyper-priors. Each plate shows a different version of DIVERSE that depends on the manner of factorization and the integrated side information. In particular, black colored nodes denote the matrices, green nodes represent the parameters of the matrices, blue ones are prior to the projection matrices, and the red nodes represent the hyper-priors of the model. See text for more details. }
\label{fig:graph}
\end{figure*}

\subsubsection{Importance weights}

We learn the importance of each data to investigate the contribution to the prediction task. In this way, we can ensure that no single side data source will dominate the prediction task, and the method will find a solution that better fits all data sets. The importances are learned by modifying the likelihood functions of HMF to include a set of importance weights $w_r, w_d, w_p, w_g$ and learning them from data. After adding the weights, the likelihoods are,
\begin{align} 
\mathbf{R} \sim \mathcal{N}(\mathbf{R};\mathbf{U} \cdot \mathbf{ S_{r}} \cdot \mathbf{V}^{\mathsf{T}})^{w_{r}}, \quad \mathbf{D} \sim  \mathcal{N}(\mathbf{D; U \cdot S_{d} \cdot U}^{\mathsf{T}})^{w_{d}} \notag\\
\mathbf{P} \sim \mathcal{N}(\mathbf{P};\mathbf{V} \cdot \mathbf{ S_{p}} \cdot \mathbf{H}^{\mathsf{T}})^{w_{p}}, \quad \mathbf{G} \sim  \mathcal{N}(\mathbf{G; H \cdot S_{g} \cdot H}^{\mathsf{T}})^{w_{g}}.
\label{eq:importance}
\end{align}

 \begin{figure*}[!t]
\centering
\includegraphics[width=\textwidth]{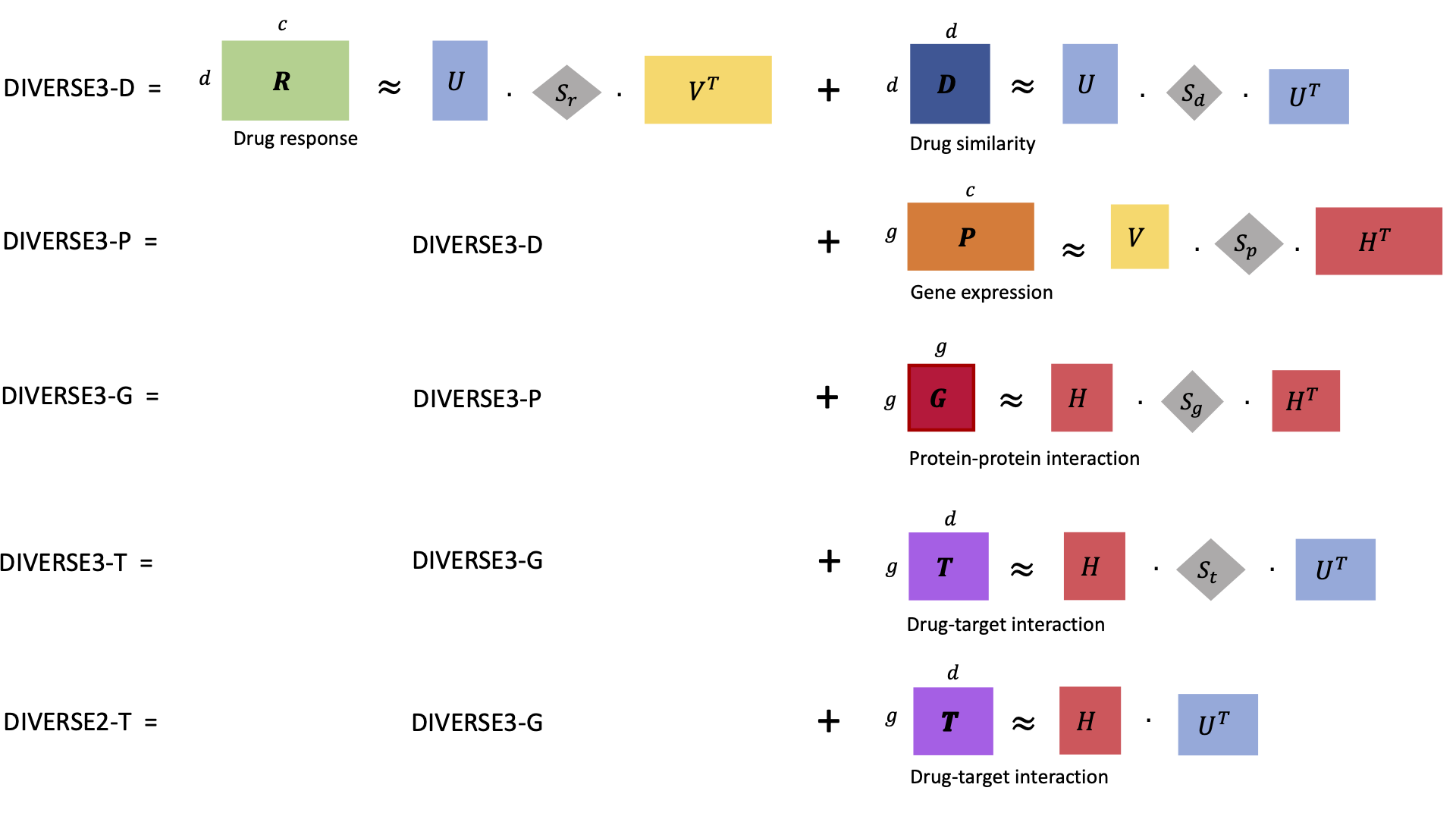}
\caption{Overview of our systematic framework, DIVERSE, of integrating multiple data sets: importance weight tri-(or bi-)matrix factorization. We start with adding $\mathbf {D}$ to $\mathbf{R}$ (first row: DIVERSE3-D). We then add $\mathbf{P}$ to DIVERSE3-D (second row: DIVERSE3-P). Similarly we add $\mathbf{G}$ to DIVERSE3-P (third row: DIVERSE3-G) and $\mathbf{T}$ to DIVERSE3-G (fourth row: DIVERSE3-T). Another option of the last addition is bi-matrix factorization, and this is the last row: DIVERSE2-T.}
\label{fig:all_matrices}
\end{figure*}

\subsection{Integrating side information} 
\label{sec:Integrating side information}

We propose a data integration framework to improve the efficiency of the prediction of anticancer drug responses in cell lines by incorporating heterogeneous data about observed relationships among cell lines, drugs, and genes. Our framework incorporates those multiple relationships as matrices into the data blocks of HMF. The model assumes that the drug responses and all side information sources are conditionally independent given the parameters so that the likelihood can be written as the product over these sources. In particular, to examine the importance of each data matrix, we integrate multiple matrices in a step-wise manner which, starting with $\mathbf{R}$, adds one data matrix one by one, in the order of $\mathbf{D}$, $\mathbf{P}$, $\mathbf{G}$ and $\mathbf{T}$. At each step, we examine the importance of the added data matrix by checking the predictive performance. The detailed relation between the datasets and their parameters, priors and hyper-priors can be seen in Fig. \ref{fig:graph} as a graphical illustration for a better understanding.

\subsubsection{Data integration: step-wise methods}
\label{sec:comparison methods}

Our framework is a step-wise workflow of gradually integrating multiple data matrices, where at each step we explore the importance of each added matrix through the importance weight, and each step is based on matrix tri-factorization (MTF) or matrix bi-factorization (MF) of HMF. We call the method DIVERSE (for {\it Bayesian Data IntegratiVE learning for drug ResponSE prediction}), particularly DIVERSE3 ({\it Importance Weight matrix-Tri-Factorization}) or DIVERSE2 ({\it Importance Weight matrix-Bi-Factorization}), depending on the manner of factorization. Fig. \ref{fig:all_matrices} shows a schematic picture of our framework, which for five data auxiliary data matrices produces five prediction methods having progressively more auxiliary data.

\begin{enumerate}[wide=0pt, listparindent=0.7cm, parsep=0pt,labelsep=5pt]

\vspace{5mm}

\item \textbf{Incorporating drug similarity data (DIVERSE3-D)}

\noindent
Drug similarity is one of the most commonly used side information sources to improve drug response prediction. We start by adding $\mathbf{D}$ to $\mathbf{R}$ to demonstrate the method yields comparable results to earlier methods using the same two data sources. For given likelihood functions of drug response and drug similarity matrices (\ref{eq:importance}), we integrate the data matrices in the total likelihood function as follows,
\begin{align}
    p(\theta|\mathbf{R},\mathbf{D}) \propto p(\theta) p(\mathbf{R}|\mathbf{U},\mathbf{S_{r}}, \mathbf{V},\tau_r)^{w_{r}}  p(\mathbf{D}|\mathbf{U},\mathbf{S_{d}},\tau_d)^{w_{d}}. 
\end{align}{}
\noindent
where $\theta$ denotes all parameters, $p(\theta)$ is the prior, and the two last terms the likelihoods for the two data sources, weighted by data set specific weights $w_r$ and $w_d$.

\vspace{5mm}

\item \textbf{Incorporating gene expression data (DIVERSE3-P)}

\noindent
Gene expression has also been utilized for the prediction of drug responses since a considerable amount of gene expression data has become publicly available. When combining gene expression data $\mathbf{P}$ with DIVERSE3-D, we can write the posterior probability as
\setlength{\arraycolsep}{0.0em}
\begin{eqnarray}
    p(\theta|\mathbf{R},\mathbf{D},\mathbf{P}) \propto & p(\theta)  p(\mathbf{R}|\mathbf{U},\mathbf{S_{r}}, \mathbf{V},\tau_r)^{w_{r}}  p(\mathbf{D}|\mathbf{U},\mathbf{S_{d}},\tau_d)^{w_{d}} \nonumber \\
    & p(\mathbf{P}|\mathbf{V},\mathbf{S_{p}},\mathbf{H},\tau_p)^{w_{p}}.
\end{eqnarray}
\setlength{\arraycolsep}{5pt}

\vspace{5mm}

\item \textbf{Incorporating protein-protein interaction data (DIVERSE3-G)}

\noindent
Protein-protein interaction is another significant source that researchers have recently started to incorporate to predicting cell line-drug associations. We integrate this information into DIVERSE3-P. The posterior distribution of the four data sets is
\setlength{\arraycolsep}{0.0em}
\begin{eqnarray}
    p(\theta|\mathbf{R},\mathbf{D},\mathbf{P},\mathbf{G}) \propto & p(\theta)  p(\mathbf{R}|\mathbf{U},\mathbf{S_{r}}, \mathbf{V},\tau_r)^{w_{r}}  p(\mathbf{D}|\mathbf{U},\mathbf{S_{d}},\tau_d)^{w_{d}} \nonumber \\
    & p(\mathbf{P}|\mathbf{V},\mathbf{S_{p}},\mathbf{H},\tau_p)^{w_{p}}p(\mathbf{G}|\mathbf{H},\mathbf{S_{g}},\tau_g)^{w_{g}}. \nonumber\\*  
\end{eqnarray}
\setlength{\arraycolsep}{5pt}

\item \textbf{Incorporating drug-target interaction data (DIVERSE3/2-T)}

\noindent
In the last step, for a given drug-target interaction data set, we have two different scenarios that decompose $\mathbf{T}$ in different ways
\setlength{\arraycolsep}{0.0em}
\begin{eqnarray}
\mathbf{T} \approx  \mathbf{H S_{t} U}^{\mathsf{T}} \quad \textrm{or} \quad  \mathbf{T} \approx  \mathbf{H  U}^{\mathsf{T}}, \quad \textrm{such that} \quad \mathbf{U},\mathbf{H},\mathbf{ S_{t}} \in \mathbb{R}^{+}. \nonumber \\ 
\label{eq:T}
\end{eqnarray}
\setlength{\arraycolsep}{5pt}
The idea behind these two distinct ways is that so far we have repeatedly used MTF, and then now we can try two cases: we 1) keep using MTF, or 2) switch to MF, where latent factors $\mathbf{H}$ and $\mathbf{U}$ can be more regularized by $\mathbf{T}$ than MTF, which might be useful for prediction. This different experimental setup will reveal the flexibility of the model. In other words, by switching to MF, which has fewer parameters, the decomposition can be more regularized.

\vspace{0.3cm}

\begin{enumerate}[wide=0pt, listparindent=1.25em, parsep=0pt,labelsep=5pt]

\item \textbf{DIVERSE3-T}

\noindent
 We decompose $\mathbf{T}$ into three matrices so that this factorization will have an advantage of using interactions between the two latent vector spaces. Especially because this data is binary, simultaneous factorization would be preferable. In this scenario, blocks are given as $\mathcal{M} = \{ \mathbf{R}, \mathbf{P}, \mathbf{T} \}$ and $\mathcal{S} = \{ \mathbf{D}, \mathbf{G} \}$.  We modify the posterior such as,
 \setlength{\arraycolsep}{0.0em}
\begin{eqnarray}
    p(\theta|\mathbf{R},\mathbf{D},\mathbf{P},\mathbf{G}) \propto & p(\theta)  p(\mathbf{R}|\mathbf{U},\mathbf{S_{r}}, \mathbf{V},\tau_r)^{w_{r}}  p(\mathbf{D}|\mathbf{U},\mathbf{S_{d}},\tau_d)^{w_{d}} \nonumber \\
    & p(\mathbf{P}|\mathbf{V},\mathbf{S_{p}},\mathbf{H},\tau_p)^{w_{p}}p(\mathbf{G}|\mathbf{H},\mathbf{S_{g}},\tau_g)^{w_{g}} \nonumber\\*  
    & p(\mathbf{T}|\mathbf{H},\mathbf{S_{t}},\mathbf{U},\tau_t)^{w_{t}}.
\end{eqnarray}
\setlength{\arraycolsep}{5pt}

 \vspace{-4mm}
 
\item  \textbf{ DIVERSE2-T}

\noindent
In the second scenario, we use MF which requires fewer parameters for T, and uses latent factors of drug and gene entities obtained from the main block
 \setlength{\arraycolsep}{0.0em}
\begin{eqnarray}
    p(\theta|\mathbf{R},\mathbf{D},\mathbf{P},\mathbf{G}) \propto & p(\theta)  p(\mathbf{R}|\mathbf{U},\mathbf{S_{r}}, \mathbf{V},\tau_r)^{w_{r}}  p(\mathbf{D}|\mathbf{U},\mathbf{S_{d}},\tau_d)^{w_{d}} \nonumber \\
    & p(\mathbf{P}|\mathbf{V},\mathbf{S_{p}},\mathbf{H},\tau_p)^{w_{p}}p(\mathbf{G}|\mathbf{H},\mathbf{S_{g}},\tau_g)^{w_{g}} \nonumber\\*  
     & p(\mathbf{T}|\mathbf{H},\mathbf{U},\tau_g)^{w_{t}}.
\end{eqnarray}
\setlength{\arraycolsep}{5pt}
\end{enumerate}
\end{enumerate}

\section{Experimental evaluation}

\subsection{Data}
\label{sec:data}
We used five publicly available data sources, which consist of measurements on three types of entities: drugs, cell lines, and genes, for predicting the response of cancer cell lines. Fig. \ref{fig:all_data} shows a conceptual scheme on the relation between data sets. All data sets vary in different ranges, such as drug-target interaction is binary data set, while drug similarity data ranges between [0,100]. For the consistency between integrated data sources, we scaled all data to the range between [0,1]. The statistics of the five data sets we used in our experiments are summarized in Table \ref{Tab:data}.

\begin{table}[H]
\renewcommand{\arraystretch}{1.3}
\caption{Statistics on Five Data Sets in Our Experiments}
\label{Tab:data}
\centering
\addtolength{\tabcolsep}{-1pt}
\begin{tabular}{@{}lccccccccc@{}}
\hline
&\#drugs & \#cell lines & \#genes& observed & sources\\ \hline
$\mathbf{R}$ &  255 & 956 &  -  & 0.82  & GDSC\\
$\mathbf{D}$ &  255 & -   &  -  & 0.92  & Pubchem\\
$\mathbf{P}$ &  -   & 956 & 232 & 1     & GDSC\\
$\mathbf{G}$ &  -   & -   & 232 & 0.5   & STRING\\
$\mathbf{T}$ &  255 & -   & 232 & 0.009 & GDSC + Chembl\\
\hline
\end{tabular}
\vspace{0.2cm}

     Note: $\mathbf{R}$: drug response, $\mathbf{D}$: drug similarity, $\mathbf{P}$: gene expression, $\mathbf{G}$: protein-protein interaction, $\mathbf{T}$: drug-target interaction

\end{table}%

\subsubsection{Drug response ($\mathbf{R}$)}

We obtained drug response data ($\mathbf{R}$) from Genomics of Drug Sensitivity in Cancer (GDSC) \cite{yang2012genomics}, consisting of IC50 values that measuring drug activity concentration required for $50\%$ inhibition (a lower value of IC50 indicates a better sensitivity of a cell line to a given drug). For 265 drugs and 992 cell lines, the data had been log-transformed. After the pre-processing, we obtained 255 drugs and 956 cell lines.

\subsubsection{Drug similarity ($\mathbf{D}$)}
Drug similarity ($\mathbf{D}$), based on the chemical structural similarity between compounds, is usually used to identify compounds sharing similar biological or chemical activity. We used the PubChem Score Matrix Service \cite{kim2015pubchem} for extracting 2D similarity scores of the 255 drug compounds in $\mathbf{R}$.

\subsubsection{Gene expression ($\mathbf{P}$)}
We used gene expression data ($\mathbf{P}$) provided by the GDSC project. The data had been measured with Affymetrix Human Genome U219 Arrays, and normalized by using RMA. We used the genes found both in drug-target interactions and gene expression, resulting in 232 genes with their interactions with 956 cell lines.

\subsubsection{Drug-target interaction ($\mathbf{T}$)}
Drug–target interaction data ($\mathbf{T}$) were collected from GDSC and Chembl \cite{gaulton2017chembl} databases. We extracted drug-target interactions for 255 drugs, which also exist in the drug response matrix and 232 genes, which are found in the gene expression data.

\subsubsection{Protein-protein interaction ($\mathbf{G}$)}
PPI ($\mathbf{G}$) is very noisy but might be helpful to understand the behavior of drug responses, since drug effects can be affected by proteins through various networks, such as metabolic pathways. We retrieved protein interactions from STRING \cite{szklarczyk2010string}, which includes physical and functional associations.

\subsection{Experiment settings}
\label{sec:Experimental settings}

\subsubsection{In-matrix and out-of-matrix prediction}
We empirically evaluated the predictive performance of the five methods presented in Section \ref{sec:comparison methods} by associations between cell lines and drugs.
%by incorporating chemical and genomic information sources. 
For comparison of these methods, we considered two tasks: 

\vspace{1mm}
\noindent 
(i) {\it in-matrix prediction}: we predict missing values in $\mathbf{R}$. \ \\ 
(ii) {\it out-of-matrix prediction}: we predict values for entirely unseen drug response vectors to given cell lines.
\vspace{1mm}

\subsubsection{Determining importance weight}
An important and hard problem is to find the best set of importance values, particularly for the method with a larger number of data sets.
For this problem, we examined various number of importance values for both in-matrix and out-of-matrix predictions. We used the values in the range of [0,1] with a fixed interval, resulting in that all values are [0.2, 0.4, 0.6, 0.8, 1]. Our way of exploring a set of optimal importance values is a greedy manner. That is, we repeated the following three steps: 1) at one method in Section \ref{sec:comparison methods}, we added one data set and then tested each of all five possible importance values, meaning that we conducted five experiments for each method. 2) we adopted the value, which provides the best performance among the five possibilities. 3) we then moved to the next method to add another data set.

\subsubsection{Cross-validation}
We conducted five times 5-fold (5x5-fold) cross-validation with different random cross-validation folds where we held a subset of drugs as a test set and trained the model on the rest of the drugs. We predicted the response values for drugs in the test set, by using the trained model.

\subsubsection{Performance evaluation measures}
We evaluated the predictive performance of all methods by using the Spearman correlation coefficient (Sc) and MSE between the observed (true) and the predicted IC50s. We focused on drug-averaged Spearman correlation scores across test drugs since the correlation over all drug responses between true and predicted drug sensitivity scores might overestimate the predictive performance.

\subsection{Comparison methods}

We note that our framework allows to integrate five different data sets, while so far to the best of our knowledge, there are no bioinformatics methods, which can incorporate the five different data sets (three entity types) for prediction of drug response values. Thus we were unable to find any competing methods, which can use same data sets as our framework. 

We first used two baselines, by following \cite{ammad2014integrative}, i.e. the mean of training drug response values as a prediction for the unobserved drug responses, where we considered 1) cell-line specific mean (cls-mean) and 2) overall mean (all-mean). 
We then used two state-of-the-art machine learning approaches: MultiNMF \cite{fujita2018biomarker} and KRR \cite{murphy2012machine}. Lastly, we used DrugCellNet, which is a straightforward but efficient network interpolation method for drug response prediction. We chose DrugCellNet since DrugCellNet already outperformed standard machine learning methods such as ElasticNet, random forest and support vector regression in 
\cite{zhang2015predicting}. We note that these three methods (MultiNMF, KRR and DrugCellNet) cannot use the entire five data sets in our experiment, though DIVERSE can handle all these five. Instead, these three methods used only drug response ($\mathbf{R}$) and drug similarity ($\mathbf{D}$) data sets.

We tuned relevant hyper-parameters of MultiNMF and KRR by using grid search on each training set. We selected the size of dimensionality K of MultiNMF as 10. We followed the the same experimental procedure as described above for the compared methods, i.e. 5x5-fold cross-validation.

\subsection{Performance results}

We entirely compared the following ten methods: cls-mean (cell-line specific mean), all-mean (overall mean); MultiNMF, KRR, DrugCellNet (drug response and similarity), DIVERSE3-D, DIVERSE3-P, DIVERSE3-G and finally DIVERSE2-T and DIVERSE3-T considering two cases: i) in-matrix prediction and ii) out-of-matrix prediction.
Note that, our primary motivation is to focus on out-of matrix prediction, since this problem is more challenging, and also a more realistic setting, in which unknown drugs are given.

\begin{figure*}[!t]
\hspace{-10mm}
(a) MSE (DIVERSE3-T) \hspace{10mm}
(b) Sc  (DIVERSE3-T) \hspace{15mm}
(c) MSE (DIVERSE2-T)\hspace{10mm}
(d) Sc (DIVERSE2-T) \hspace{4mm}
    \centering
    \includegraphics[width=1\textwidth]{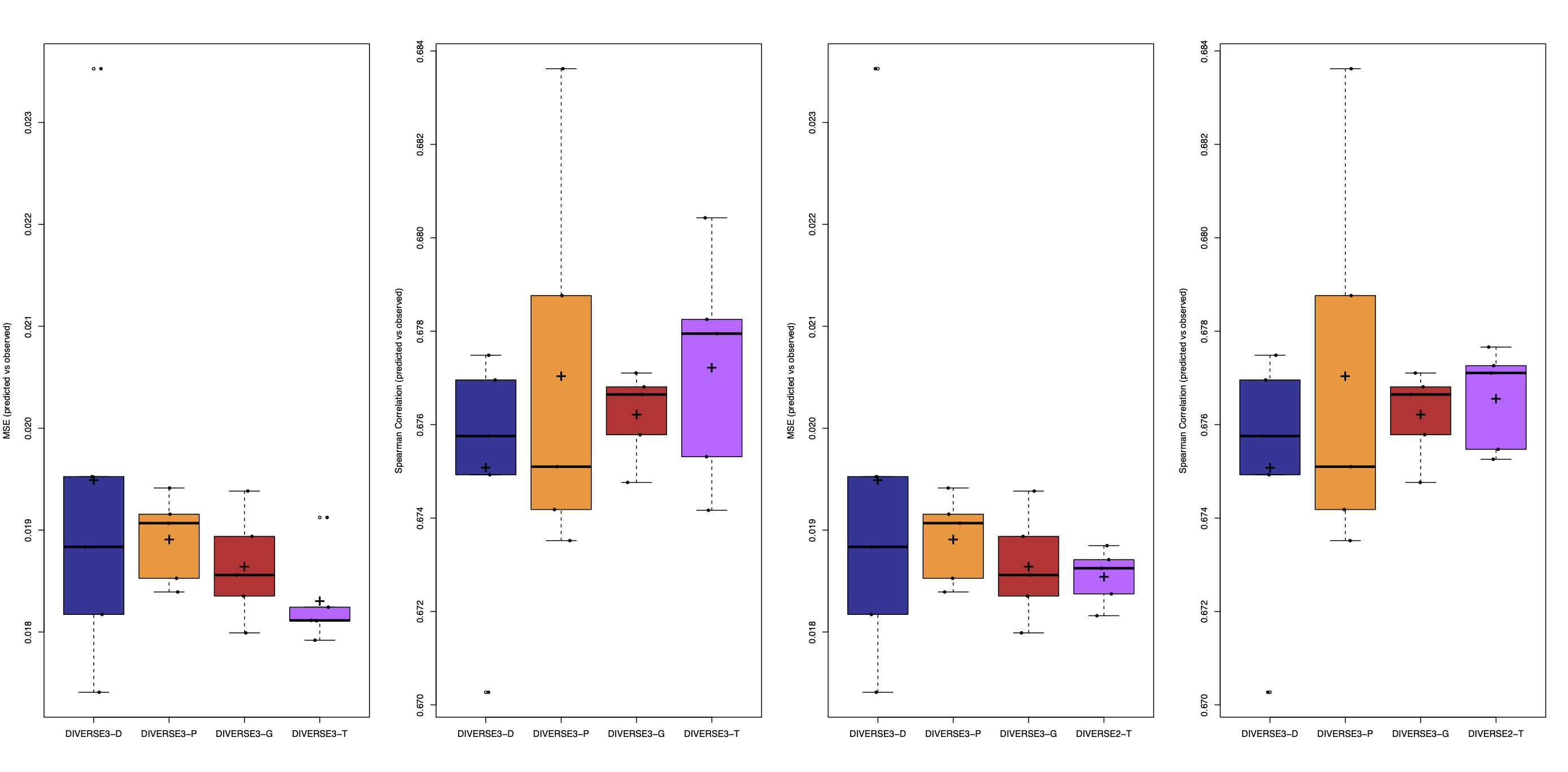}
    \caption{Performance results of out-of-matrix prediction under 5x5-fold cross validation. (a) MSE and (b) Sc of the case of DIVERSE3-T when $\mathbf{T}$ is added, and (c) MSE and (d) Sc of the case of DIVERSE2-T when $\mathbf{T}$ is added. The thick black horizontal line in each box is the median, while black cross of each box is the mean. Thus please use each black cross to examine the results.}
    \label{fig:out-matrix-results}
\end{figure*}

\subsubsection{In-matrix prediction performance}

\begin{table}[!t]
\renewcommand{\arraystretch}{1.3}
\caption{MSE and Sc (average scores of 5x5 cross-validation) of Ten Compared Methods in In-matrix Prediction}
\label{Tab:in-matrix}
\centering
\begin{tabular}{@{}lccccccccccc@{}}\hline
 &   MSE $\pm$ Std. Dev.                 &    Sc $\pm$ Std. Dev.   \\
 \hline
cls-mean         &   $0.5226 \pm 0.0036$         &     --                  \\
all-mean         &   $0.4181 \pm 0.0726$         &     --                  \\
MultiNMF         &   $0.0209 \pm 0.0020$         &     $0.4717 \pm 0.0041$ \\
KRR              &   $0.0625 \pm 0.0046$         &     $0.0111 \pm 0.0250$ \\
DrugCellNet      &   $0.0532 \pm 0.0002$         &     $0.4481 \pm 0.0056$ \\
DIVERSE3-D           &   $0.0047 \pm 0.0001$         &     $0.4852 \pm 0.0023$ \\
DIVERSE3-P           &   $0.0048 \pm 0.0008$         &     $0.4833 \pm 0.0040$ \\
DIVERSE3-G           &   $0.0047 \pm 0.0001$         &     $0.4840 \pm 0.0039$ \\
DIVERSE2-T           &   $0.0047 \pm 0.0001$         &     $0.4841 \pm 0.0037$ \\
DIVERSE3-T           &   $0.0048 \pm 0.0001$         &     $0.4844 \pm 0.0032$ 
\\\hline
\end{tabular}
\vspace{0.2cm}

Note: Std. Dev. stands for standard deviation. 
\end{table}

Table \ref{Tab:in-matrix} shows the MSE and Sc of the ten compared methods under 5x5-fold cross-validation. The five methods of DIVERSE achieved significantly smaller MSE than the two baselines and three state-of-the-art methods. The performance of MultiNMF was the second (after DIVERSE) in both MSE and Sc, which might be reasonable, because MultiNMF is also based on NMF, though using drug responses and drug similarity data only. The performance of KRR was worse compared to MultiNMF and DrugCellNet, probably because regression may have negative output values, although all true outputs are here known to be non-negative. Overall, DIVERSE outperformed all compared methods, indicating that DIVERSE would be most robust to predict missing data. On the other hand, the differences between the five methods of DIVERSE were rather unclear. This result implies that adding multiple data sets might not necessarily be so effective for in-matrix prediction.

\subsubsection{Out-of-matrix prediction performance}

%The method which achieved the best MSE and Sc is in bold.

Table~\ref{Tab:out-matrix} shows the MSE and Sc of the ten compared methods under 5x5-fold cross-validation, where the lowest MSE and largest Sc are highlighted in bold. The five methods of DIVERSE again achieved significantly smaller MSE and higher Sc scores than the other five methods. MultiNMF was worst among the existing methods, implying that NMF is ineffective for out-of-matrix prediction though being useful for filling missing values. DrugCellNet was next to DIVERSE in both MSE and Sc. Among the five methods of DIVERSE, starting with DIVERSE3-D, the MSE was decreasing like DIVERSE3-P, DIVERSE3-G, finally resulting in DIVERSE3-T, the smallest value among all ten compared methods. This result indicates that step-wise data set addition of DIVERSE worked well for integrating heterogeneous data sets. Also, this result was confirmed by Sc, where Sc was basically increased by adding more data sets, finally reaching 0.6772, which was again the highest among all ten compared methods.

\begin{table}[!t]
\renewcommand{\arraystretch}{1.3}
\caption{MSE and Sc (average scores of 5x5 cross-validation) of Ten Compared Methods in Out-of-matrix Prediction}
\centering
\label{Tab:out-matrix}
\begin{tabular}{@{}lccc@{}}\hline
  &   MSE $\pm$ Std. Dev.                 &    Sc $\pm$ Std. Dev.   \\\hline
cls-mean         &   $0.5227 \pm 0.0027$         &     --                   \\
all-mean         &   $0.4181 \pm 0.0726$         &     --                   \\
MultiNMF         &   $0.1581 \pm 0.0721$         &     $0.1457 \pm 0.0180$    \\
KRR              &   $0.0764 \pm 0.0125$         &     $0.2976 \pm 0.0361$     \\
DrugCellNet      &   $0.0455 \pm 0.0044$         &     $0.3423 \pm 0.0259$    \\
DIVERSE3-D           &   $0.0194 \pm 0.0049$         &     $0.6750 \pm  0.0186$  \\
DIVERSE3-P           &   $0.0189 \pm  0.0049$        &     $0.6770 \pm  0.0188$\\
DIVERSE3-G           &   $0.0186 \pm  0.0035$        &     $0.6762  \pm  0.0179$\\
DIVERSE2-T           &   $0.0185 \pm 0.0040$         &      $0.6765 \pm 0.0187$ \\
DIVERSE3-T           &   $\boldsymbol{0.0183 \pm 0.0033}$        &   $\boldsymbol{0.6772 \pm 0.0193}$ 
\\ \hline
\end{tabular}
\vspace{0.2cm}

     Note: Std. Dev. stands for standard deviation. 
\end{table}

We can compare our two ways of adding $\mathbf{T}$, i.e. DIVERSE3-T and DIVERSE2-T, from a performance perspective. Table~\ref{Tab:out-matrix} shows the MSE of DIVERSE3-T was 0.0183, which was smaller than the MSE of DIVERSE2-T which was 0.185. This difference sounds very slight, and we checked box plots of MSE and Sc, which are shown in Fig. \ref{fig:out-matrix-results}, where the left two figures ((a) MSE and (b) Sc) show the case of DIVERSE3-T (DIVERSE3-D $\rightarrow$ DIVERSE3-P $\rightarrow$ DIVERSE3-G $\rightarrow$ DIVERSE3-T) and the right two figures ((c) MSE and (d) Sc) show the case of DIVERSE2-T (DIVERSE3-D $\rightarrow$ DIVERSE3-P $\rightarrow$ DIVERSE3-G $\rightarrow$ DIVERSE2-T). In these figures, the thick black line in each box shows the median, and the mean value of each case is shown by black cross. We observe that the mean value of (a) decreased clearly (particularly at the last DIVERSE3-T), while the decrease of the mean of (c) is rather mild, particularly at the last DIVERSE2-T. Thus we can see DIVERSE3-T achieved a better performance than DIVERSE2-T,\enlargethispage{\baselineskip}  implying that matrix tri-factorization is better than matrix bi-factorization here. Eventually, strong regularization by $\mathbf{T}$ (see the bottom of Fig. ~\ref{fig:all_matrices}) might not be so useful.

\begin{table}[!t]
\renewcommand{\arraystretch}{1.3}
\caption{Importance Weights of the Best Performance Case of Each Time in 5x5-fold Cross-validation}
\centering
\label{Tab:importance_weights}
\begin{tabular}{@{}lccccccccr@{}}\hline
 & 1st & 2nd & 3rd & 4th & 5th & Average \\\hline
 DIVERSE3-D & 0.2 & 0.2 & 0.2 & 0.2 & 0.2 & 0.2\\
 DIVERSE3-P & 0.8 & 0.8 & 0.2 & 0.8 & 1 & 0.68\\
 DIVERSE3-G & 1 & 0.2 & 0.4 & 0.6 & 0.8 & 0.6\\
 DIVERSE2-T & 0.4 & 0.4 & 0.6 & 0.4 & 1 & 0.56 \\
 DIVERSE3-T & 1 & 1 & 0.8 & 0.4 & 1 & 0.84 
\\\hline
\end{tabular}
\end{table}

Finally, we examine the importance weights, which were computed when a data set is newly added in the greedy procedure of DIVERSE. Table \ref{Tab:importance_weights} shows the importance weights obtained by the best performance case when we added a newly data set (for example $\mathbf{D}$ for DIVERESE3-D) in DIVERSE,
for each of the five times 5-fold cross-validation, and also the average over the five times.
The highest average importance weight was obtained by DIVERSE3-T, indicating that the importance weight was large when $\mathbf{T}$, i.e. drug-target interactions, was added. However, other data sets also had rather large average importance weights, like 0.56 to 0.68, except $\mathbf{D}$, i.e. drug similarity, with always 0.2.
Interestingly, this result implies that drug similarity might not have been so significant.

\subsubsection{Case study}

To find potential drugs for our cancer cell lines from a different perspective, we checked how well data integration worked for prediction improvement for individual cases. The idea here is if we have drugs for which prediction was improved by integrating more data, we can predict whether the drug is useful for a given cell line. From this assessment, we could raise three sample drugs: CUDC-101, Gemcitabine, and SN-38 (known as also Irinotecan), for which primary targets are EGFR/ERBB2, pyrimidine antimetabolite, and TOP1 respectively. Table~\ref{Tab:case_study} shows how each version of DIVERSE improved the correlation score between the observed and predicted values of each of the three drugs. These results indicate the existence of highly predictable drugs, and also, as a methodology, our framework of data integration worked on predicting the IC50 values. Furthermore, these results show that integrating biological side information is useful to predict unseen drugs from existing drug screening values and improve efficiency.
\begin{table}[H]
\renewcommand{\arraystretch}{1.3}
\caption{Average MSE and Spearman Correlation Scores Over 5x5-fold Cross-validation for Multiple Cancer Cell Lines}
\centering
\addtolength{\tabcolsep}{-1pt}
\label{Tab:case_study}
\begin{tabular}{@{}lccccccc@{}}\hline
& \multicolumn{2}{c}{CUDC101}
& \multicolumn{2}{c}{Gemcitabine}
& \multicolumn{2}{c}{SN-38} \\
& MSE & Sc
& MSE & Sc
& MSE & Sc \\ \hline
DIVERSE3-D 
& 0.00096 & 0.916
& 0.00182 & 0.930
& 0.00146 & 0.861 \\
DIVERSE3-P 
& 0.00099 & 0.910
& 0.00203 & 0.923
& 0.00122 & 0.884 \\
DIVERSE3-G 
& 0.00092 & 0.915
& 0.00211 & 0.919
& 0.00165 & 0.841 \\
DIVERSE2-T 
& 0.00087 & 0.922
& 0.00189 & 0.929
& \textbf{0.00119} & \textbf{0.894} \\
DIVERSE3-T 
& \textbf{0.00081} & \textbf{0.926}
& \textbf{0.00138} & \textbf{0.948}
& 0.00121 & 0.887 \\ \hline
\end{tabular}
\end{table}

\begin{figure}[!t]
\centering
\includegraphics[width=0.6\linewidth]{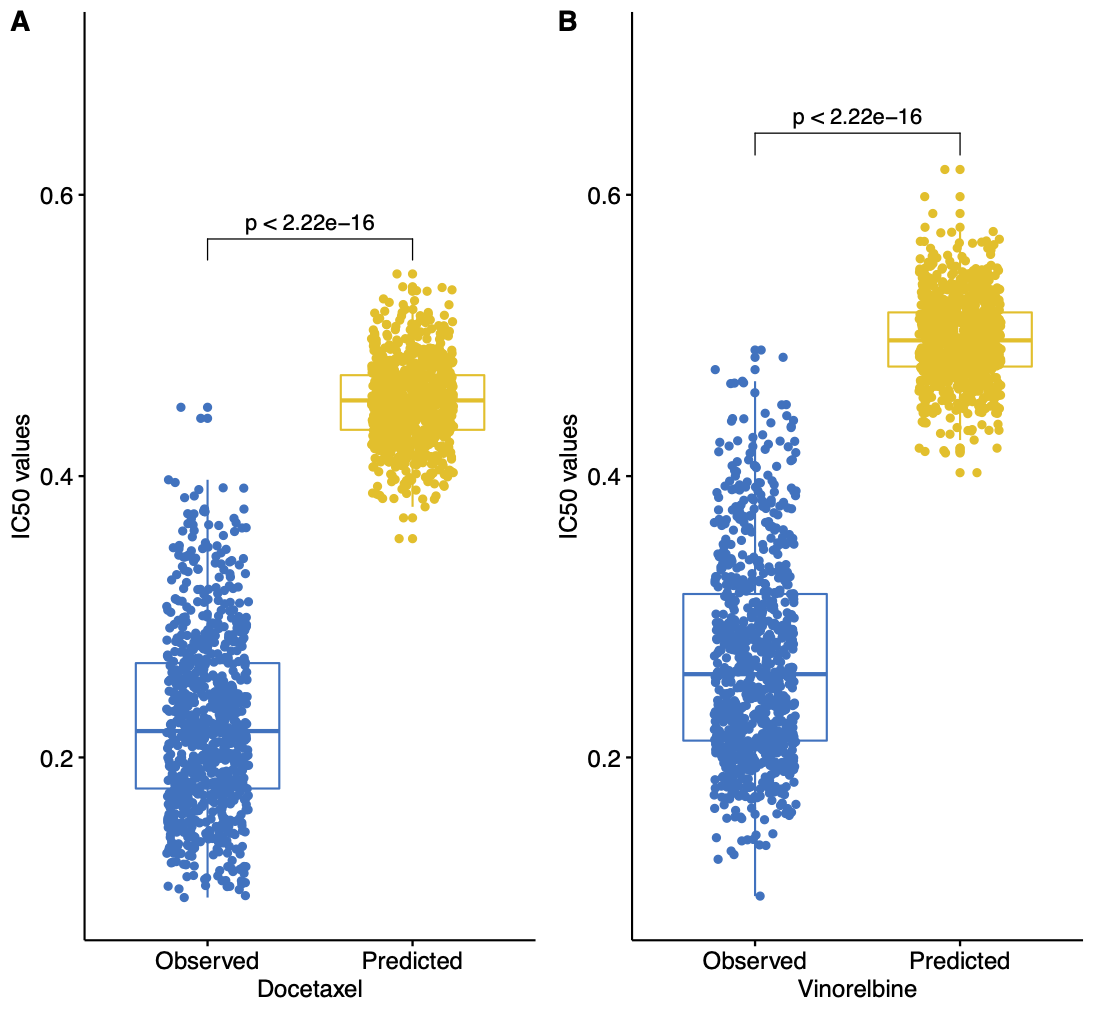}
\caption{Box plots of two test drugs from GDSC dataset: A. Docetaxel and B. Vinorelbine. The $t$-test was used to measure the statistical difference in the mean between the predicted and observed response values of cell lines for each drug.}
\label{fig:p_value}
\end{figure}

To understand the results obtained by DIVERSE3-T more, we trained DIVERSE3-T by using all available data and predicted responses of unseen chemical compounds. The idea is that we may find a new drug, if a  positive response is predicted by DIVERSE3-T, even if the observed (true) data are not positives, i.e. negatives. Thus we first chose compounds with different values between the predicted (by DIVERSE3-T) and observed values and then ran a one-sided paired $t$-test to confirm the significance of the difference in the mean. We then obtained two potential drugs for our cancer cell lines:
%three powerful drugs mainly targeting breast cancer: 
1) Docetaxel  
%, 2) Tamoxifen 
and 2) Vinorelbine.
Fig. \ref{fig:p_value} shows the distributions of predicted and observed values for these two compounds.
For these two drugs, the values predicted by DIVERSE3-T were significantly higher than the observed values, with p-values $< 2.22e-16$.
We further checked the literature and some more details of these two drugs can be given as follows:
1) Docetaxel is one of the powerful known drugs which has many interactions with other drugs and also known interactions with mitosis pathway such as TUBB1 and MAP2/MAP4 proteins; where $13\%$ of response values are missing. 
%2) Tamoxifen is a protein kinase C inhibitor targeting estrogen receptors, where most of the responses are already known with $95.08\%$ observed fraction rate. 
2) Vinorelbine is a similar drug with Docetaxel where approximately $10\%$ of the entries are missing and targeting microtubule destabilizer in the mitosis pathway. 
Thus these two might be used as cancer drugs.

\section{Discussion and conclusion}

Cancer is a complex disease affected by genotypes and associated with other factors including phenotypes, environmental exposures, drugs, and chemical molecules. No single data source can explain the underlying factors and capture complexity. Machine learning methods that combine heterogeneous data from multiple sources have thus emerged as critical, statistical and computational approaches. Although various methods have been developed for anticancer drug response prediction, challenges remain in many aspects, such as choosing the informative data sources suitable for training and testing models, computational approaches that can incorporate many sources efficiently, and deciding how such models are evaluated and validated.

We have proposed DIVERSE, Bayesian matrix factorization with importance weights, a new framework to infer drug-cell line associations incorporating side information collected from different sources. To overcome the challenges which we mentioned above, we focused on integrating as much data as possible, which reaches five data sets in our experiments. In DIVERSE, the data can be systematically integrated in a cascade manner, examining the importance of each incorporated data set. We empirically validated the performance of  DIVERSE, comparing with five other methods, including three state-of-the-art methods, under 5x5-fold cross-validation. Experimental results indicate that DIVERSE would clearly be useful for out-of-matrix prediction, which is a real-world setting and much harder than in-matrix prediction. In particular, the results indicate that the performance (MSE) of DIVERSE was smoothly improved by the step-wise addition of more data sets. These advantages of DIVERSE were confirmed by several case studies. Even though our proposed framework has achieved encouraging results, it cannot avoid the following disadvantages. We prefer using Gibbs sampling since it is one of the most robust optimization methods which gives the ability to estimate the full Bayesian posterior, especially for sparse data sets. However, it converges slowly and requires additional samples to estimate the posterior compared to other optimization techniques such as variational Bayes. Secondly, incorporating more meaningful data effectively could improve the predictive performance such as the drug similarity data which is based on the 2D chemical structural similarity between compounds in our work. Even though 2D features give sufficient features to represent a drug, 3D structure features might also play a crucial role. Similar cases also can be considered for other integrated sources. For example; better identification of gene-drug associations provides a comprehensive understanding of effective treatments for patients \cite{chen2019hogmmnc}. We benefit from one-gene-to-one-drug associations for drug-gene interaction data, however, if different drugs interact with each other and targets ``many-genes-to-many drugs'', the drug response therapy may be further improved \cite{huang2020evaluation,cai2018identifying}.

Predictive performance might be further improved if more informative data sources can be incorporated into our methods, and exploring new data sets would be direct future work. Another direction might be to predict the response of drug combinations by integrating more data since drug combination therapy could provide an effective strategy to overcome drug resistance \cite{chen2016nllss}. There might be another challenge rising here because of the big data problems that require carefully chosen feature selection methods. Also in machine learning, various techniques, including those in deep learning, are continuously being developed. Incorporating such new techniques into our method for better prediction or interpretability would be interesting future work.

\bibliographystyle{unsrt}  
%\bibliography{main.bbl}  %%% Remove comment to use the external .bib file (using bibtex).

\end{document}